\documentclass[twocolumn,showpacs,preprintnumbers,amsmath,amssymb,prb,superscriptaddress,aps,10pt]{revtex4-1}

\usepackage{graphicx}

\begin{document}

\title{Charge transfer statistics of transport through Majorana bound states}
\author{H. Soller}
\affiliation{Institut f\"ur Theoretische Physik,
Ruprecht-Karls-Universit\"at Heidelberg,\\
 Philosophenweg 19, D-69120 Heidelberg, Germany}
\author{A. Komnik}
\affiliation{Institut f\"ur Theoretische Physik,
Ruprecht-Karls-Universit\"at Heidelberg,\\
 Philosophenweg 19, D-69120 Heidelberg, Germany}

\date{\today}

\begin{abstract}
We analyse the full counting statistics of charge transfer through a Majorana bound state coupled to an STM tip and show how they can be used for an unambiguous identification of the bound state at the end of the wire. Additionally, we show how to generate Majorana bound states in a simple setup involving a ferromagnetic wire on a superconducting substrate. 
\end{abstract}


\maketitle

\section{Introduction}
Majorana Bound States (MBSs) at the ends of quantum wires, which represent a solid state realisation of Majorana fermions, have been proposed as a fault-tolerant quantum memory in a topological quantum computer \cite{1063-7869-44-10S-S29}. However, their unambiguous detection is difficult since they are neutral quasiparticles. Depending on the scheme used to generate the MBS several proposals have been discussed. Most studies focus on conductance properties \cite{PhysRevB.82.180516,PhysRevLett.103.237001,PhysRevB.81.184525} or the AC Josephson effect \cite{kwon,PhysRevB.79.161408}. However, so far noise properties of systems with a MBS have attracted little attention: the noise has been investigated in systems of coupled vortex cores \cite{PhysRevLett.98.237002} and the cross correlation of currents through coupled MBSs at the ends of a quantum wire has been used to probe nonlocality \cite{PhysRevLett.101.120403}. Furthermore noise properties of an isolated Majorana fermion coupled to a normal metal has been discussed in \cite{PhysRevB.83.153415}. A discussion of noise properties of transport through a MBS in a superconductor at the end of a quantum wire so far is missing but is essential not just for detection but for a complete understanding of the readout schemes for quantum memory that have been developed so far \cite{2011arXiv1107.4338L,PhysRevLett.106.090503}. We close this gap by not just calculating the noise but directly the full counting statistics (FCS), which represent the ultimate low-frequency characteristics of transport \cite{levitov-1996-37}. We use a generic model for the Majorana fermion and compare our results to a specific model for an InAs nanowire based setup.\\
In the second part of this paper we discuss a possible way to obtain the MBSs. The original scheme  for realizing the unpaired Majorana fermions involved topological superconductors \cite{PhysRevLett.100.096407,PhysRevB.61.10267} but materials such as $\mbox{Sr}_2\mbox{Ru}\mbox{O}_4$ are hard to handle \cite{PhysRevLett.93.167004}. Therefore manifold schemes have been developed to overcome the need for nanostructured topological superconductors: one could use the coupling of vortices in a bulk $p$-wave superconductor \cite{PhysRevLett.98.237002} or fermionic cold atoms \cite{PhysRevLett.98.010506} or it is possible to mimic a $p$-wave superconducting wire by using spin-orbit coupling in combination with the proximity effect \cite{PhysRevLett.106.057001,oppen,1367-2630-13-5-053016,2011arXiv1103.2746B}. First experiments in this direction have already been done \cite{2012arXiv1202.2323W,nature}. One may also use spin-active scattering to generate $p$-wave superconductivity in a ferromagnetic halfmetal wire \cite{PhysRevB.83.054513,PhysRevB.84.060510}, however also relying on materials that require special care \cite{PhysRevB.83.054513}. In the second part of this work we will show how to overcome this requirement only using a ferromagnet (FM) and a conventional superconductor (SC).

\section{Detection of Majorana Fermions}

\begin{figure}[ht]
\centering
\includegraphics[width=7cm]{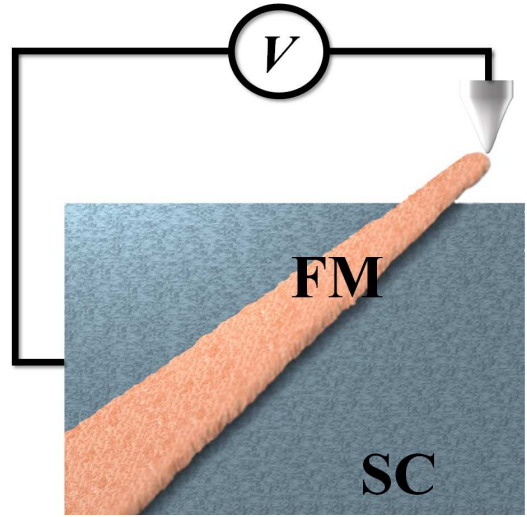}
\caption{Setup discussed in this work: a superconductor (SC) is tunnel coupled to a ferromagnetic wire (FM). Due to the induced triplet superconductivity the wire hosts a MBS at the end whose transport characteristics are investigated when coupled to an STM tip. The coupling between the SC and the FM is much better than the one between the MBS and the STM tip so that a voltage ($V$) drop is assumed only to occur between these two parts of the system.}
\label{fig1}
\end{figure}
We consider a generic setup for investigating a Majorana fermion. Using either of the abovementioned proposals for generating the MBSs the effective low-energy behavior of the system is that of a $p$-wave SC \cite{PhysRevLett.105.177002,PhysRevLett.107.036801,PhysRevLett.105.077001}. Therefore the MBS in the wire coupling to the STM tip may be treated as an effective $p$-wave SC with the Hamiltonian \cite{PhysRevLett.92.127001,PhysRevLett.107.036801}
\begin{eqnarray}
H_{\mathrm{eff}} &=& \sum_{k} \epsilon_k \Psi_{k,\uparrow}^+ \Psi_{k,\uparrow} \nonumber\\
&& + \sum_k (\Delta_p \Psi_{k,s,\uparrow}^+ \Psi_{-k,s,\uparrow}^+ + \Delta_p^*\Psi_{-k,s,\uparrow} \Psi_{k,s,\uparrow}), \label{eff}
\end{eqnarray}
where $\Delta_p$ refers to the effective (temperature-dependent) $p$-wave gap and we use units $e = \hbar = k_B =1$. Consequently we start from Kitaev's model \cite{1063-7869-44-10S-S29}. Compared to an $s$-wave SC (see also Eq. (\ref{bcs})) we break time-reversal symmetry and spin-rotation symmetry. Electron-hole symmetry at the Fermi level remains and we have the typical MBS situation \cite{Bocquet2000628,PhysRevB.59.13221,PhysRevB.62.8249,PhysRevB.63.224204,PhysRevB.71.245124}.\\
In this formulation the wire is spin polarized with a DOS $\rho_0(1+P)\Delta_p/\sqrt{\Delta_p^2 - \omega^2}$ in the wide band limit. $P$ refers to a possible spin polarisation.\\
Considering a FM as the host material of the MBS (a spinless fermion) may be created in a FM by inducing $p$-wave correlations via spin-active scattering as will be shown below. The mechanism for creating spin-active scattering at the interface between the FM and the SC is the ferromagnetic exchange field both in the bulk and the interface of the FM. The magnetic moment in the interface and the bulk may be misaligned either due to a thin FM layer, spin-orbit coupling or magnetic anisotropy \cite{PhysRevB.81.094508}.\\
Since the Hamiltonian in Eq. (\ref{eff}) is very similar to a $s$-wave SC Hamiltonian the FCS will share several properties with previous results on the FCS with $s$-wave SCs \cite{PhysRevB.50.3982,fdots}. The relevant differences to such previous results are the nature of the SC ($p$-wave vs. $s$-wave) and the presence of the Majorana fermion which leads to an observable phase shift in the reflection coefficient.\\
The tip is described as a normal metal with a flat band density of states $\rho_{0T}$ using electron field operators $\Psi_{T,k,\sigma}$
\begin{eqnarray*}
H_{\mathrm{STM}} = \sum_{k,\sigma} \epsilon_k \Psi_{T,k,\sigma}^+ \Psi_{T,k,\sigma}.
\end{eqnarray*}
The STM tip is held at chemical potential $\mu_T$ whereas the wire in Eq. (\ref{eff}) is held at chemical potential $\mu=0$ in accordance with previous studies of SC contacts \cite{PhysRevB.50.3982}. The description of tunneling between the wire in Eq. (\ref{eff}) and the STM tip is given by the usual tunneling Hamiltonian \cite{PhysRevLett.8.316,PhysRev.163.341}. We explicitely include the possibility of additional phase shifts $\tilde{\phi}$ during the Andreev reflection \cite{2010arXiv1012.3867H,PhysRevB.70.134510}. Such phase shifts may either vary depending on the spin due to spin-active scattering \cite{2010arXiv1012.3867H,PhysRevB.70.134510} or depending on the particle being an electron or a hole due to the presence of Majorana fermions \cite{PhysRevLett.106.057001}. In contrast to normal conducting tunnel contacts such phase shifts cannot be gauged away since Andreev reflection is a coherent two-electron process such that these phase shifts appear in physical results
\begin{eqnarray}
H_{T,\mathrm{STM}} &=& \gamma_{T} [e^{i\tilde{\phi}/2}\Psi_\uparrow^+(x=0) \Psi_{k,T,\uparrow}(x=0) + \mathrm{h.c.}]. \label{tunnel}
\end{eqnarray}
This way the additional phase shift $\tilde{\phi}$ accounts for the topological phase of the system. For simplicity we choose $\tilde{\phi} = \pm \phi$ for electrons/holes. Since we want to describe the FCS of charge transfer through the MBS we need to calculate the probability distribution function $P(Q)$ of transferring $Q$ units of charge during a given (long) measurement time $\tau$. Physical observables can then be calculated as averages with respect to this distribution function. However, instead of directly calculating $P(Q)$ it is often more convenient to calculate its cumulant generating function (CGF) $\ln \chi(\lambda) = \ln \sum_Q e^{i \lambda Q} P(Q)$. This allows for a calculation using Keldysh Green's functions \cite{PhysRevB.70.115305,PhysRevB.73.195301} via the fundamental expression
\begin{eqnarray}
\chi(\lambda) = \langle T_{\cal C} e^{-i \int_{\cal C} dt T^{\lambda} (t}\rangle_0, \label{eq3}
\end{eqnarray}
where $T^{\lambda}(t)$ in Eq. (\ref{eq3}) refers to the tunneling Hamiltonian in Eq. (\ref{tunnel}) with the substitution $\Psi_\uparrow(x=0) \rightarrow \Psi_\uparrow (x=0) e^{-i \lambda/2}$. Using the Hamiltonian approach \cite{PhysRevB.54.7366} we can calculate the CGF that will be the sum of two contributions for the different energy regimes with respect to the $p$-wave gap
\begin{eqnarray}
\ln \chi(\lambda) &=& \ln \chi_e(\lambda) + \ln \chi_A(\lambda). \label{cgf}
\end{eqnarray}
The system in question undergoes a phase transition to a topologically non-trivial phase. We use an effective description of this phase in which the only effect is the emergence of a Majorana bound state at the interface which contributes a scattering phase shift of $\phi = \pi$ to the scattering matrix of our system \cite{PhysRevLett.106.057001}. Consequently the two Andreev bound states at $\pm \Delta_{p}$ that are present in the topologically trivial phase \cite{PhysRevB.54.7366} merge (also in non-equilibrium) into one bound state at \cite{PhysRevB.38.8823,2012arXiv1203.2643C}
\begin{eqnarray*}
\epsilon_{\mathrm{MBS}} = \pm \Delta_{p} \cos (\pi/2) = 0
\end{eqnarray*}
This zero-energy bound state affects the properties of Andreev reflection in our system. The chemical potential in this topologically non-trivial phase is zero as chosen before. The non-trivial phase would be characterised by a negative chemical potential \cite{1063-7869-44-10S-S29,PhysRevB.61.10267}.\\
The effect of such phase shifts on Andreev reflection have been discussed before for unconventional superconductors \cite{PhysRevLett.74.3451}. In this case the superconducting gap has a strong dependence on the angle of the resulting contact between the normal metal and the unconventional ($d$-wave) superconductor. It is interesting to note that the angle plays a role which is related to the phase shift mentioned above and leads to similar effects in the conductance.\\
Concerning the FCS of charge transfer above the gap we may, however, neglect Andreev reflection and branch crossing due to the small tunnel coupling of the STM tip and arrive at a Levitov-Lesovik formula \cite{levitov-1996-37} with a transmission coefficient from Tinkhams semiconductor model \cite{tinkham} in accordance with previous works on normal-SC contacts \cite{PhysRevB.50.3982}
\begin{eqnarray*}
&& \ln \chi_{e}(\lambda) = \tau \int \frac{d\omega}{2\pi} \ln \{ 1 + T_e(\omega) [n_{T} (1- n_F) (e^{i \lambda} -1) \nonumber\\
&& + n_F(1- n_T) (e^{-i \lambda}-1)]\}\theta\left(\frac{\omega - \Delta_{p}}{\Delta_{p}}\right),
\end{eqnarray*}
where $n_T$ and $n_F$ refer to the Fermi distributions of the tip and the FM respectively. $T_e(\omega) = 4 \Gamma  |\omega| /\sqrt{\omega^2 - \Delta_{p}^2}$ using $\Gamma = (1 + P) \pi^2 \rho_{0T} \rho_0 \gamma_{T}^2$.\\
Finally we get to the contribution below the gap that allows to observe the topologically non-trivial phase
\begin{eqnarray}
\ln \chi_A(\lambda) &=& \tau \int \frac{d\omega}{2\pi} \ln \{1+ T_A(\omega) [(e^{2i\lambda} -1) n_T(1-n_{T+}) \nonumber\\
&& + (e^{-2i\lambda} -1) n_{T+} (1-n_T)]\} \theta\left(\frac{\Delta_{p} - \omega}{\Delta_{p}}\right), \nonumber\\
T_A(\omega) &=& \frac{T^2}{1+ R^2 - 2R \cos(2 \arccos (\omega/\Delta_{p}) + \pi)}, \label{ta}
\end{eqnarray}
where $T = 4 \Gamma$, $R= 1- T$ and $n_{T+} =  1- n_{T}(-\omega)$ refers to the Fermi distribution of the holes. The additional phase factor $\pi$ in Eq. (\ref{ta}) is the difference to the topologically trivial case where only the Andreev reflection phase shift $\arccos (\omega/\Delta_{p})$ would be present \cite{2010arXiv1012.3867H}.\\
Deriving $P(Q)$ from the above expressions can be done numerically but in the limiting cases $V\gg \Delta_p$ (see Eq. (\ref{Pe})) and $V\ll \Delta_p$ (see Eq. (\ref{Pqmbs})) we can even derive them in closed form
\begin{eqnarray}
P_e(Q) &=& \left(\begin{array}{c} M \\ Q \end{array}\right) (4\Gamma)^{(Q)} (1- 4\Gamma)^{M - (Q)} \label{Pe}\\
P_A(2Q) &=& \left(\begin{array}{c} M \\ Q \end{array}\right), \; P_A(2Q+1) = 0,  \label{Pqmbs}
\end{eqnarray}
where $M= \tau V/\pi$. Eq. (\ref{Pqmbs}) reflects that at voltages $V \ll \Delta_p$ transport through the MBS is perfect but only occurs in pairs (only even numbers of charges are possible), whereas Eq. (\ref{Pe}) shows that voltages $V\gg \Delta_p$ transport proceeds via single-electron transfer.\\
We compare the results of this generic model with model calculations based on a Hamiltonian which describes an InAs nanowire on an Al or Nb substrate. The model parameters will be chosen such that we get as close as possible to a single channel situation as treated above in order to have a comparable situation. The Bogoliubov-de Gennes Hamiltonian in this specific situation is
\begin{eqnarray}
{\cal H}_{\mathrm{BdG}} &=& \left(\begin{array}{cc} H_R - E_F & \Delta \sigma_y \\ \Delta^* \sigma_y & E_F -H_R^* \end{array}\right),
\end{eqnarray}
which couples electron and hole excitations near the Fermi level $E_F$ through an $s$-wave superconducting order parameter $\Delta$.\\
The excitations in this model are confined to a wire of width $W$ in the $x$-$y$-plane of the semiconductor surface inversion layer, where their dynamics is governed by the Rashba Hamiltonian
\begin{eqnarray*}
H_R &=& \frac{\bf{p}^2}{2 m_{\mathrm{eff}}} + U(\bf{r}) + \frac{\alpha_{so}}{\hbar} (\sigma_x p_y -\sigma_y p_x) \nonumber\\
&& + \frac{1}{2} g_{\mathrm{eff}} \mu_B B \sigma_x.
\end{eqnarray*}
The spin is coupled to the momentum $\bf{p} = -i \hbar \partial /\partial \bf{r}$ by the Rashba effect, and polarized through the Zeeman effect by a magnetic field $B$ parallel to the wire (in $x$-direction). Characteristic length and energy scales are $l_{so} = \hbar^2 /m_{\mathrm{eff}} \alpha_{so}$ and $E_{so} = m_{\mathrm{eff}} \alpha_{so}^2 /\hbar^2$ that we choose as units for our model calculations. Typical values in InAs are $l_{so} = 100$nm, $E_{so} = 0.1$meV, $E_Z = \frac{1}{2} g_{\mathrm{eff}} \mu_B B = 1$meV at $B=1$T.\\
The electrostatic potential $U= U_{\mathrm{barrier}} + \delta U$ is the sum of a gate potential $U_{\mathrm{barrier}}$ and a possible impurity potential $\delta U$ that may vary randomly from site to site. We consistently checked that the properties of our system remain stable upon not too strong impurity potentials so that we will neglect their effect in the following.\\
The simulation is done on a lattice with $N_W = 39$ sites in the transversal direction, corresponding to a width $W= (N_W +1) a$, where the lattice constant $a$ was chosen such that both the spin-orbit length and the transversal wave functions of the wire are properly resolved, $l_{so} = Na$ ($N = 40$ was chosen). We checked that this choice of parameters corrresponds to a typical single-channel situation.\\
Now we may study the consequences of the additional phase factor in Eq. (\ref{ta}). We obtain a perfectly transmitting channel at $\omega =0$ which corresponds to a zero-bias conductance of $2e^2/h$ in SI units at zero temperature that can be calculated from the first derivative of the CGF and agrees with previous studies \cite{PhysRevB.82.180516,PhysRevLett.103.237001,PhysRevB.81.184525}
\begin{figure*}[ht]
\centering
\includegraphics[width=16cm]{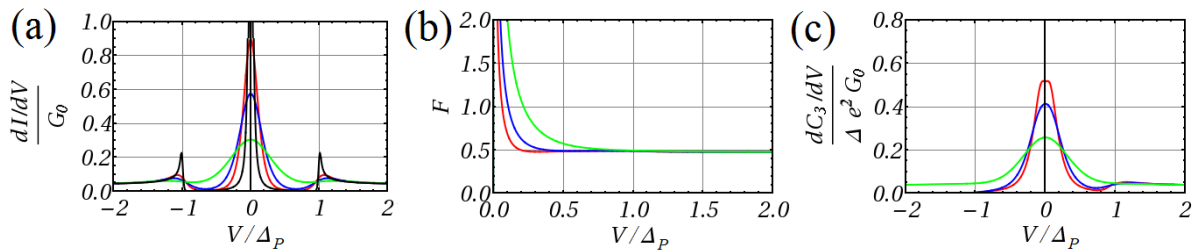}
\caption{Results derived from the CGF in Eq. (\ref{cgf}) for the conductance, Fano factor and differential third cumulant for $\Gamma = 0.01$ and $T=0.06\Delta$ (red), $T=0.1\Delta$ (blue) and $T=0.2\Delta$ (green): in (a) we show the result for the conductance. The black curve is the result for $T=0.01\Delta$, where we observe that the conductance reaches (almost) $G_0$. Furthermore we see the proximity induced SC correlations at $V= \pm \Delta_p$. The Fano in (b) at $V=0$ shows only thermal noise and then quickly goes to $0.5$ describing charge transport of a single spin species. Part (c) shows the differential third cumulant that follows the form of the conductance in (a).}
\label{fig2}
\end{figure*}\\
In Fig. \ref{fig2} we plotted the results for the conductance, Fano factor $F= S/(2 \cdot I)$ and the differential third cumulant that one may obtain by taking the first, second and third derivative with respect to $\lambda$ of Eq. (\ref{cgf}). The figure shows the results for three different temperatures. We see that while observing the SC correlations in the conductance (Fig. \ref{fig2} (a)) really requires to go to low temperatures with respect to the proximity induced gap, the features caused by the MBS remain stable upon varying temperature. Consequently everything that is crucial, is to induce a sizeable $p$-wave gap in the wire.\\
Concerning noise we see that, as the MBS opens a perfectly transmitting channel, we only observe thermal noise at $V=0$. For finite voltage the Fano factor quickly approaches $0.5$ referring to the spin-polarized $p$-wave SC. The differential third cumulant follows a Levitov-Reznikov-like behavior \cite{PhysRevB.70.115305} since it follows the form of the conductance $dC_3/dV \propto dI/dV$.\\
We check our results by comparison also to the numerical model for a single channel described above. In the case of  an InAs nanowire we calculate the transmission and reflection amplitudes from the model described above which directly give access to the conductance \cite{PhysRevB.46.12841}. A typical result is shown in Fig. \ref{fig:condscan} where we consider a high tunnel barrier in order to connect with the results for a small interface transparency shown above. We observe that the Majorana bound state and the sidepeaks due to the $p$-wave gap are clearly resolved and coincide with the calculations from the generic model. At higher energies the band structure in InAs starts to play a role and deviations are expected.
\begin{figure}[ht]
\centering
\includegraphics[width=8cm]{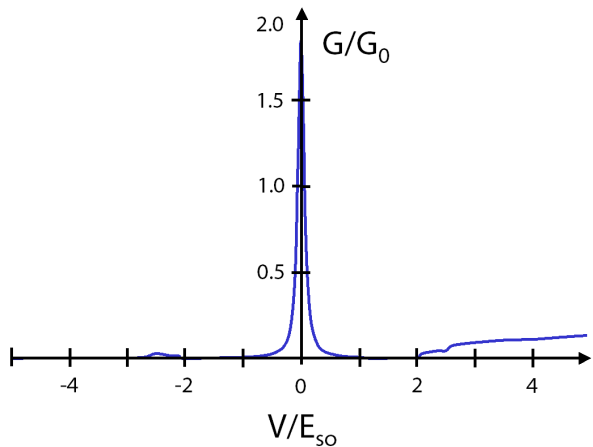}
\caption{Conductance for the situation with a high tunnel barrier. We consider a very thin wire $W = l_{so}/5$,with spin-orbit coupling, a Zeeman term along the wire, $E_Z = 6 E_{so}$ at Fermi energy $E_F = 123 E_{so}$ with a lattice fulfilling $a = l_{so}/100$. The gap is $\Delta = 4E_{so}$. The barrier is $W_b = 2a$ wide and has a potential height (on-site potential energy) of $U_b = 600E_{so}$.}
\label{fig:condscan}
\end{figure}\\
From the transmission and reflection amplitudes we can also calculate the noise at $T=0$ via \cite{PhysRevB.50.3982}
\begin{eqnarray}
\frac{dS}{dV} &=& G_0 e \mathrm{Tr} \left[r_{eh}^+ r_{eh} + r_{ee}^+ r_{ee} +2 r_{ee}^+ r_{ee} r_{eh}^+ r_{eh} \right. \nonumber\\
&& \left. - (r_{eh}^+ r_{eh})^2 - (r_{ee}^+ r_{ee})^2\right],
\end{eqnarray}
and use for the Fano factor
\begin{eqnarray}
F &=& \frac{\int_0^{eV} dS/dV d\omega}{2 \int_0^{eV} dI/dV d\omega}.
\end{eqnarray}
The result for the Fano factor is shown in Fig. \ref{fig:fano} for the same tunnel barrier strength as above. We observe again that the Fano factor quickly approaches 0.5 up to a small disagreement since the tunnel barrier is not infinite. However, the Fano factor starts at zero and not at infinity. This is due to the presence of finite temperature in the calculations for the generic model. At zero temperature the Fano factor has to be zero since the Majorana creates a perfectly transmitting channel.
\begin{figure}[ht]
\centering
\includegraphics[width=8cm]{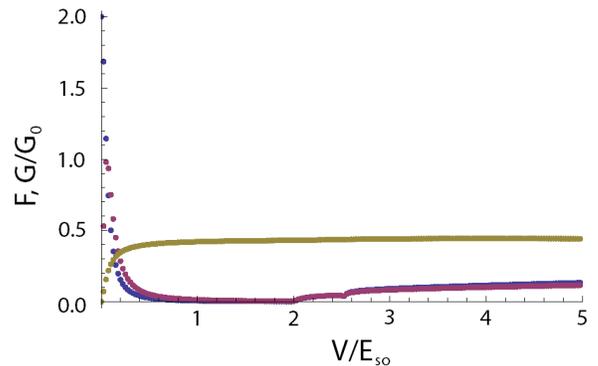}
\caption{Differential conductance (blue), differential noise(purple) and Fano factor (yellowish) using the same parameters as in Fig. \ref{fig:condscan}.}
\label{fig:fano}
\end{figure}\\
For zero temperature the generic model in Eq. (\ref{cgf}) produces a similar result for the Fano factor as shown in Fig. \ref{fig:fanozero}. This result is also in accordance with the study in \cite{PhysRevB.83.153415} where a Fano factor of zero at zero voltage and 0.5 at high voltages was predicted for an isolated Majorana fermion coupled to a normal metal.
\begin{figure}[ht]
\centering
\includegraphics[width=8cm]{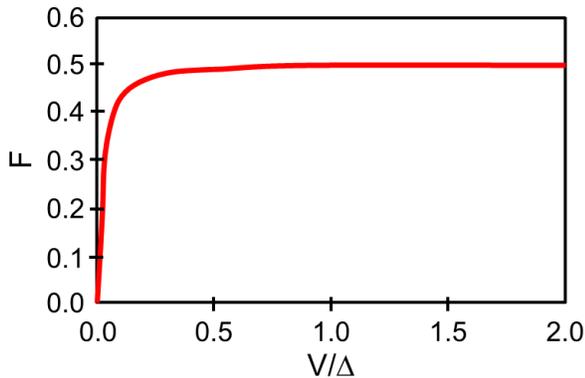}
\caption{Results derived from the CGF in Eq. (\ref{cgf}) for the Fano factor at $T=0\Delta$ and $\Gamma = 0.01$. The result has the same form as the Fano factor in Fig. \ref{fig:fano}}
\label{fig:fanozero}
\end{figure}\\
The three signatures identified: a peak in the conductance, the form of the Fano factor and the behavior of the third cumulant therefore give a very clear signature to be searched for in future experiments. States at the interface that are not bound states could show a similar pattern in the conductance but not in the Fano factor and the third cumulant. The only bound states that could show a similar behavior are Andreev bound states that, however, so far have been observed only in SC-FM heterostructures. Additionally these states would not be localised at the end of the wire. Therefore moving the tip away from the end of the wire must lead to the dissapearance of the peak in the conductance and therefore allows for an unambiguous identification of the MBS.\\
Especially, the method allows to discriminate the Majorana conductance peak from the Kondo conductance peak by taking the higher cumulants into account. A Kondo resonance could give a similar feature in the conductance but the Fano factor would be different: if one compares the system to a Kondo impurity between two normal conductors \cite{PhysRevLett.97.016602} we expect $F=10/6$ at zero bias. If one compares to the situation of a Kondo impurity between a superconductor and a normal conductor one would expect the typical doubling of shot noise \cite{Soller2011425} so that one would observe the same features as in Fig. \ref{fig2}(b) but with a doubled Fano factor and a third cumulant multiplied (as in  Fig. \ref{fig2}(c)) by four. This way the method described above allows to clearly discriminate the Majorana conductance features from the ones due to a Kondo impurity both between normal conductors and between a superconductor and a normal conductor.

\section{Majoranas in superconductor ferromagnet hybrids}

In the second part of this work we want to describe the generation of the MBS at the end of the wire proximity coupled to the SC. Since the contact between the SC and the FM is much better than the one between the FM and the STM tip we consider a voltage drop only between the latter two parts of the system. We approach the coupling of the SC to the FM wire in two steps: first, we describe a lateral tunnel contact in order to extract the general gap structure induced in the wire due to the presence of the SC. In a second step we use this description in order to write down a low-energy effective Hamiltonian describing the wire. The starting point for the first part is the tunnel contact between the SC and the FM,
\begin{eqnarray}
H = H_F + H_S + H_T. \label{hfull}
\end{eqnarray}
$H_F$ refers to the Hamiltonian of the FM that is described in the language of electron field operators $\Psi_{k,f,\sigma}$ by the Stoner model with an exchange energy $h_{ex}$ as in \cite{melin-2004-39}
\begin{eqnarray}
H_F &=& \sum_{k\|, \sigma} \epsilon_{k\|} \Psi_{k\|,f,\sigma}^+ \Psi_{k\|,f,\sigma} \nonumber\\
&&  - h_{ex} \sum_{k\|} (\Psi_{k\|,f,\uparrow}^+ \Psi_{k\|,f,\uparrow} - \Psi_{k\|,f,\downarrow}^+ \Psi_{k\|,f,\downarrow}). \label{hferro}
\end{eqnarray}
The FM represents a fermionic continuum with a spin-dependent DOS $\rho_{0,\sigma} = \rho_0 (1+ \sigma P)$. We denote the momenta by $k\|$ since we assume to have a flat wire so that only momenta parallel to the wire direction are allowed. The SC Hamiltonian $H_S$ is given by the typical BCS expression using electron field operators $\Psi_{k,s,\sigma}$ as
\begin{eqnarray}
H_S &=& \sum_{k,\sigma} \epsilon_k \Psi_{k,s,\sigma}^+ \Psi_{k,s,\sigma} \nonumber\\
&& + \Delta \sum_k (\Psi_{k,s,\uparrow}^+ \Psi_{-k,s,\downarrow}^+ + \Psi_{-k,s,\downarrow} \Psi_{k,s,\uparrow}). \label{bcs}
\end{eqnarray}
Concerning the tunneling between the SC and the FM we need to include the recently predicted \cite{PhysRevLett.101.257001,PhysRevB.77.064517} and observed \cite{2010arXiv1012.3867H,PhysRevB.83.081305,fdots} spin-activity of the interface in SC-FM hybrids.  Manifold effects like spin-orbit coupling, magnetic anisotropy or spin relaxation may give rise to spin-activity of interfaces \cite{PhysRevB.81.094508}. Previous studies of SC-FM contacts used a scattering states description \cite{PhysRevB.81.094508,PhysRevLett.101.257001,PhysRevB.77.064517,PhysRevLett.102.227005} in order to introduce a spin-active scattering angle as a phenomenological parameter to characterise the interface or a wave-function matching technique \cite{PhysRevB.83.054513}. We follow \cite{PhysRevB.84.060510} by introducing a second spin-active tunneling term in $H_T$
\begin{eqnarray}
H_T &=& \sum_{k,\sigma} \gamma_{1} [\Psi_{k\|,f,\sigma}^+ \Psi_{k,s,\sigma} + \mathrm{h.c.}],\nonumber\\
&& + \sum_{k,\sigma} \gamma_{2} [\Psi_{k\|,f,\sigma}^+ \Psi_{k,s,-\sigma} + \mathrm{h.c.}]. \label{tunnelwire}
\end{eqnarray}
Such description assumes that spin-active scattering can be effectively described by two tunneling Hamiltonians: one which flips the spin of scattered electron and one which does not, whereas in the description in \cite{PhysRevB.81.094508} the electrons acquire a phase shift depending on their spin. The equivalence of both descriptions has been discussed in \cite{fdots}, where it was shown that in typical superconductor-ferromagnet hybrids the position of the Andreev bound states can be used to determine the spin-mixing angle. Both descriptions are effective descriptions of spin-active scattering capturing the microscopic specifics of the interface.\\
As in \cite{PhysRevB.84.060510} the above model can be solved exactly as it is quadratic in fermion fields.\\
We need to obtain a $p$-wave gap leading to MBSs at the wire ends. We proceed via diagonalization of the Hamiltonian in Eq. (\ref{hfull}) which is possible via considering a 1D SC coupled to a 1D FM and averaging over the additional momenta $k\bot$ in the SC afterwards. In this case our model Hamiltonian in the 1D case can be written as
\begin{eqnarray}
H_{1D} &=& \frac{1}{2} \sum_{k\|} \Psi_{k\|}^+ H_{k\|}^{BdG} \Psi_{k\|}, \; \mbox{where}\\
H_{k\|}^{BdG} &=& \left[\begin{array}{cc} h^{FM} & T_{k\|} \\ T_{k\|}^+ & \Lambda_{BdG} \end{array}\right],
\end{eqnarray}
where $\Psi_{k\|} = (\Psi_{k\|,f,\uparrow}, \Psi_{k\|,f,\downarrow},\Psi_{-k\|,f,\uparrow}^+,\Psi_{-k\|,f,\downarrow}^+,$ $\gamma_{k\|, \uparrow}, \gamma_{k\|, \downarrow}, \gamma_{-k\|, \uparrow}^+,\gamma_{-k\|, \downarrow}^+)$ and $h^{FM}$ refers to Eq. (\ref{hferro}) written in terms of the introduced $\Psi_{k\|}$ fields. $T_{k\|}$ and $\Lambda_{BdG}$ refer to Eqs. (\ref{tunnelwire}) and (\ref{bcs}) written in terms of $\Psi_{k\|}$ fields where $\gamma_{k\|,\sigma}$ operators refer to Bogoliubov quasiparticle annihilation operators.\\
We disregard possible effects of disorder in accordance with previous experimental studies \cite{PhysRevB.85.180512}. The effects of disorder have been previously studied in \cite{PhysRevLett.106.057001}.\\
The Hermitian matrix $H_{k\|}^{BdG}$ can now be diagonalized $U_{k\|}^+ H_{k\|}^{BdG} U_{k\|}$ with the unitary matrix $U_{k\|}$. This procedure in \cite{PhysRevB.81.094508} allows to express the pairing amplitude in terms of elements of $U_{k\|}$
\begin{eqnarray*}
\langle \Psi_{-k\|,f,\uparrow} \Psi_{k\|,f,\uparrow}\rangle &=& (U_{k\|})_{33}^* (U_{k\|})_{13} + (U_{k\|})_{34}^* (U_{k\|})_{14} \nonumber\\
&& + (U_{k\|})_{37}^* (U_{k\|})_{17} + (U_{k\|})_{38}^* (U_{k\|})_{18}.
\end{eqnarray*}
Averaging this expression over $k\bot$ leads to a momentum dependent pairing gap $\Delta_{\uparrow\uparrow}(k)$. However, the general solution is quite complicated. Since, for the low energy behavior relevant for the presence of a Majorana fermion, only low momenta are relevant, we can consider the case $k=0$ and obtain the following result for $h_{ex} = 0$ and $\gamma_1 = \gamma_2 \neq 0$
\begin{eqnarray}
\frac{\Delta_{\uparrow\uparrow}}{\Delta} = \frac{2(1+ \gamma^2) - 2 \sqrt{1 + 2\gamma^2}}{\gamma^2}.
\end{eqnarray}
In this case for $h_{ex} = 0$ of course $\Delta_{\uparrow\uparrow} = \Delta_{\downarrow\downarrow}$. For finite $h_{ex}$ the two proximity induced gaps are different.\\
The additional $p$-wave correlations $\langle \Psi_{\uparrow,\downarrow}^+ \Psi_{\uparrow,\downarrow}^+ \rangle$ can be traced back to the spin-activity of the interface \cite{RevModPhys.77.935}. However, due to the exchange splitting of the FM $s$-wave correlations and $p$-wave correlations of the type $1/\sqrt{2} (|\uparrow\downarrow\rangle + |\downarrow \uparrow\rangle)$ will decay on a length scale $\hbar/(p_{F\uparrow} - p_{F\downarrow})$ so that only the equal spin correlations will remain \cite{PhysRevB.81.094508}. The $p$-wave correlations $\langle \Psi_{\uparrow,\downarrow}^+ \Psi_{\uparrow,\downarrow}^+ \rangle$ will penetrate into the FM on a length scale $\hbar v_{F,\uparrow, \downarrow}/\Delta$. Therefore we assume the wire thickness $d$ to be $\hbar/(p_{F\uparrow} - p_{F\downarrow}) \ll d < \hbar v_{F,\uparrow, \downarrow}/\Delta$.

Since the equal-spin $p$-wave correlations only involve one spin species their order parameter may coexist with ferromagnetic correlations. The unequal-spin $p$-wave and $s$-wave correlations do not have this property and cannot coexist with ferromagnetic correlations.\\
This consideration generalizes the result obtained in \cite{PhysRevB.83.054513,PhysRevB.81.094508}: in the case of a ferromagnetic halfmetal no spin-$\downarrow$ exists and only one $p$-wave gap survives, which leads to the emergence of MBSs. However, for a finite exchange field in the FM always both spin species will be present.\\
In this case we might still obtain a Majorana fermion but via a mechanism more akin to the one in InAs/InSb nanowires \cite{PhysRevLett.105.177002,PhysRevLett.107.036801,PhysRevLett.105.077001}, where one uses spin-orbit coupling in combination with an applied magnetic field. In our proposal we can use spin-active scattering and ferromagnetism in the same way: spin-active scattering is introduced as in Eq. (\ref{tunnelwire}) incorporating the spin-active scattering at the interface via the additional tunnel Hamiltonian and we use the ferromagnetic wire as a source for an exchange field. The spin-activity of the interface will give rise to $p$-wave correlations in the wire and the exchange field will select a preferred spin-direction so that a quasi spin-polarized $p$-wave superconductor will emerge that allows to host the Majorana fermion.\\
The simplest Hamiltonian describing the FM wire in the presence of the SC reads following our effective description above and incorporating the effect of spin-active scattering as the emergence of a $p$-wave gap \cite{PhysRevLett.105.177002,PhysRevLett.107.036801,PhysRevLett.105.077001,Volkov1995261}
\begin{eqnarray*}
H_{\mathrm{wire}} &=& \int dx \Psi^+(x) {\cal H} \Psi(x), \; \Psi^+ = (\Psi_\uparrow^+, \Psi_\downarrow^+, \Psi_\downarrow, \Psi_\uparrow),\\
&& \mbox{where}\\
{\cal H} &=& \left[\frac{p^2}{2m} - \mu\right] \tau_z - h_{ex} \sigma_z + \left(\begin{array}{cc} 0 & \Delta_{\uparrow\uparrow} \\ \Delta_{\downarrow\downarrow} & 0 \end{array}\right) \sigma_x.
\end{eqnarray*}
$\Psi_{\uparrow, \downarrow}(x)$ annihilates spin-$\uparrow$ ($\downarrow$) electrons at position $x$. The Pauli matrices $\tau$ and $\sigma$ operate in spin- and particle-hole space, respectively. $\mu$ is the chemical potential, that we choose to be zero.\\
The spectrum can be revealed as in \cite{PhysRevLett.105.177002,PhysRevLett.107.036801,PhysRevLett.105.077001} by squaring ${\cal H}$ twice, which yields $E_{\sigma, \pm} = \sigma h_{ex} \pm \sqrt{\Delta_{\sigma\sigma}^2+ \xi_p^2}$, where $\xi_p = p^2/(2m)$. As in the aforementioned works the gap $E_0$ near $p=0$ is the key to the emergence of MBSs. We find
\begin{eqnarray}
E_0 = h_{ex} - \Delta_{\uparrow\uparrow}.
\end{eqnarray}
In the proposal using spin-orbit coupled wires two regimes corresponding to the topologically trivial and nontrivial case exist depending on the relative strength of the magnetic field compared to the proximity induced gap. In our case $h_{ex} \gg \Delta_{\uparrow\uparrow,\downarrow\downarrow}$ for typical ferromagnets \cite{PhysRevB.25.527}. Therefore we always obtain an exchange field-dominated (or strong interaction induced) gap and hence the wire will be in its topological phase. The end of the wire can now be characterised by a sharp drop of the chemical potential, which closes the gap. Since this transition corresponds to a transition out of the topological phase MBSs will be localized at the wire ends.

\section{Conclusions}

In conclusion we have presented the first results for FCS of charge transfer through the generated MBS and discussed how the results can be used for an unambiguous identification of the MBS in future experiments. Additionally we demonstrated that just using a SC and a FM for the MBS generation is sufficient if using the newly discovered spin-active scattering effects of the interface in SC-FM heterostructures. \\
The authors would like to thank J. Dahlhaus, B. Braunecker, S. Maier and L. Hofstetter for many interesting discussions. The financial support was provided by the Center of Quantum Dynamics of the University of Heidelberg.


\begin{thebibliography}{58}
\expandafter\ifx\csname natexlab\endcsname\relax\def\natexlab#1{#1}\fi
\providecommand{\url}[1]{\texttt{#1}}
\providecommand{\href}[2]{#2}
\providecommand{\path}[1]{#1}
\providecommand{\DOIprefix}{doi:}
\providecommand{\ArXivprefix}{arXiv:}
\providecommand{\URLprefix}{URL: }
\providecommand{\Pubmedprefix}{pmid:}
\providecommand{\doi}[1]{\href{http://dx.doi.org/#1}{\path{#1}}}
\providecommand{\Pubmed}[1]{\href{pmid:#1}{\path{#1}}}
\providecommand{\bibinfo}[2]{#2}
\ifx\xfnm\relax \def\xfnm[#1]{\unskip,\space#1}\fi
\bibitem[{Kitaev(2001)}]{1063-7869-44-10S-S29}
\bibinfo{author}{A.~Y. Kitaev}, \bibinfo{journal}{Physics-Uspekhi}
  \bibinfo{volume}{44} (\bibinfo{year}{2001}) \bibinfo{pages}{131}.
\bibitem[{Flensberg(2010)}]{PhysRevB.82.180516}
\bibinfo{author}{K.~Flensberg}, \bibinfo{journal}{Phys. Rev. B}
  \bibinfo{volume}{82} (\bibinfo{year}{2010}) \bibinfo{pages}{180516}.
\bibitem[{Law et~al.(2009)Law, Lee, and Ng}]{PhysRevLett.103.237001}
\bibinfo{author}{K.~T. Law}, \bibinfo{author}{P.~A. Lee},
  \bibinfo{author}{T.~K. Ng}, \bibinfo{journal}{Phys. Rev. Lett.}
  \bibinfo{volume}{103} (\bibinfo{year}{2009}) \bibinfo{pages}{237001}.
\bibitem[{Linder et~al.(2010)Linder, Tanaka, Yokoyama, Sudb\o{}, and
  Nagaosa}]{PhysRevB.81.184525}
\bibinfo{author}{J.~Linder}, \bibinfo{author}{Y.~Tanaka},
  \bibinfo{author}{T.~Yokoyama}, \bibinfo{author}{A.~Sudb\o{}},
  \bibinfo{author}{N.~Nagaosa}, \bibinfo{journal}{Phys. Rev. B}
  \bibinfo{volume}{81} (\bibinfo{year}{2010}) \bibinfo{pages}{184525}.
\bibitem[{{H.-J. Kwon} et~al.(2004){H.-J. Kwon}, {K. Sengupta}, and {V.M.
  Yakovenko}}]{kwon}
\bibinfo{author}{{H.-J. Kwon}}, \bibinfo{author}{{K. Sengupta}},
  \bibinfo{author}{{V.M. Yakovenko}}, \bibinfo{journal}{Eur. Phys. J. B}
  \bibinfo{volume}{37} (\bibinfo{year}{2004}) \bibinfo{pages}{349--361}.
\bibitem[{Fu and Kane(2009)}]{PhysRevB.79.161408}
\bibinfo{author}{L.~Fu}, \bibinfo{author}{C.~L. Kane}, \bibinfo{journal}{Phys.
  Rev. B} \bibinfo{volume}{79} (\bibinfo{year}{2009}) \bibinfo{pages}{161408}.
\bibitem[{Bolech and Demler(2007)}]{PhysRevLett.98.237002}
\bibinfo{author}{C.~J. Bolech}, \bibinfo{author}{E.~Demler},
  \bibinfo{journal}{Phys. Rev. Lett.} \bibinfo{volume}{98}
  (\bibinfo{year}{2007}) \bibinfo{pages}{237002}.
\bibitem[{Nilsson et~al.(2008)Nilsson, Akhmerov, and
  Beenakker}]{PhysRevLett.101.120403}
\bibinfo{author}{J.~Nilsson}, \bibinfo{author}{A.~R. Akhmerov},
  \bibinfo{author}{C.~W.~J. Beenakker}, \bibinfo{journal}{Phys. Rev. Lett.}
  \bibinfo{volume}{101} (\bibinfo{year}{2008}) \bibinfo{pages}{120403}.
\bibitem[{Golub and Horovitz(2011)}]{PhysRevB.83.153415}
\bibinfo{author}{A.~Golub}, \bibinfo{author}{B.~Horovitz},
  \bibinfo{journal}{Phys. Rev. B} \bibinfo{volume}{83} (\bibinfo{year}{2011})
  \bibinfo{pages}{153415}. 
\bibitem[{Liu and Baranger(2011)}]{2011arXiv1107.4338L}
\bibinfo{author}{D.~E. Liu}, \bibinfo{author}{H.~U. Baranger},
  \bibinfo{journal}{Phys. Rev. B} \bibinfo{volume}{84} (\bibinfo{year}{2011})
  \bibinfo{pages}{201308}.
\bibitem[{Flensberg(2011)}]{PhysRevLett.106.090503}
\bibinfo{author}{K.~Flensberg}, \bibinfo{journal}{Phys. Rev. Lett.}
  \bibinfo{volume}{106} (\bibinfo{year}{2011}) \bibinfo{pages}{090503}.
\bibitem[{Levitov et~al.(1996)Levitov, Lee, and Lesovik}]{levitov-1996-37}
\bibinfo{author}{L.~S. Levitov}, \bibinfo{author}{H.~W. Lee},
  \bibinfo{author}{G.~B. Lesovik}, \bibinfo{journal}{J. Math. Phys.}
  \bibinfo{volume}{37} (\bibinfo{year}{1996}) \bibinfo{pages}{4845}.
\bibitem[{Fu and Kane(2008)}]{PhysRevLett.100.096407}
\bibinfo{author}{L.~Fu}, \bibinfo{author}{C.~L. Kane}, \bibinfo{journal}{Phys.
  Rev. Lett.} \bibinfo{volume}{100} (\bibinfo{year}{2008})
  \bibinfo{pages}{096407}.
\bibitem[{Read and Green(2000)}]{PhysRevB.61.10267}
\bibinfo{author}{N.~Read}, \bibinfo{author}{D.~Green}, \bibinfo{journal}{Phys.
  Rev. B} \bibinfo{volume}{61} (\bibinfo{year}{2000})
  \bibinfo{pages}{10267--10297}.
\bibitem[{Murakawa et~al.(2004)Murakawa, Ishida, Kitagawa, Mao, and
  Maeno}]{PhysRevLett.93.167004}
\bibinfo{author}{H.~Murakawa}, \bibinfo{author}{K.~Ishida},
  \bibinfo{author}{K.~Kitagawa}, \bibinfo{author}{Z.~Q. Mao},
  \bibinfo{author}{Y.~Maeno}, \bibinfo{journal}{Phys. Rev. Lett.}
  \bibinfo{volume}{93} (\bibinfo{year}{2004}) \bibinfo{pages}{167004}.
\bibitem[{Tewari et~al.(2007)Tewari, Das~Sarma, Nayak, Zhang, and
  Zoller}]{PhysRevLett.98.010506}
\bibinfo{author}{S.~Tewari}, \bibinfo{author}{S.~Das~Sarma},
  \bibinfo{author}{C.~Nayak}, \bibinfo{author}{C.~Zhang},
  \bibinfo{author}{P.~Zoller}, \bibinfo{journal}{Phys. Rev. Lett.}
  \bibinfo{volume}{98} (\bibinfo{year}{2007}) \bibinfo{pages}{010506}.
\bibitem[{Akhmerov et~al.(2011)Akhmerov, Dahlhaus, Hassler, Wimmer, and
  Beenakker}]{PhysRevLett.106.057001}
\bibinfo{author}{A.~R. Akhmerov}, \bibinfo{author}{J.~P. Dahlhaus},
  \bibinfo{author}{F.~Hassler}, \bibinfo{author}{M.~Wimmer},
  \bibinfo{author}{C.~W.~J. Beenakker}, \bibinfo{journal}{Phys. Rev. Lett.}
  \bibinfo{volume}{106} (\bibinfo{year}{2011}) \bibinfo{pages}{057001}.
\bibitem[{Alicea et~al.(2011)Alicea, Yuval, Refael, von Oppen, and
  Fisher}]{oppen}
\bibinfo{author}{J.~Alicea}, \bibinfo{author}{O.~Yuval},
  \bibinfo{author}{G.~Refael}, \bibinfo{author}{F.~von Oppen},
  \bibinfo{author}{M.~P.~A. Fisher}, \bibinfo{journal}{Nat. Phys.}
  \bibinfo{volume}{7} (\bibinfo{year}{2011}) \bibinfo{pages}{412–417}.
\bibitem[{Wimmer et~al.(2011)Wimmer, Akhmerov, Dahlhaus, and
  Beenakker}]{1367-2630-13-5-053016}
\bibinfo{author}{M.~Wimmer}, \bibinfo{author}{A.~R. Akhmerov},
  \bibinfo{author}{J.~P. Dahlhaus}, \bibinfo{author}{C.~W.~J. Beenakker},
  \bibinfo{journal}{New Journal of Physics} \bibinfo{volume}{13}
  (\bibinfo{year}{2011}) \bibinfo{pages}{053016}.
\bibitem[{Romito et~al.(2012)Romito, Alicea, Refael, and von
  Oppen}]{2011arXiv1103.2746B}
\bibinfo{author}{A.~Romito}, \bibinfo{author}{J.~Alicea},
  \bibinfo{author}{G.~Refael}, \bibinfo{author}{F.~von Oppen},
  \bibinfo{journal}{Phys. Rev. B} \bibinfo{volume}{85} (\bibinfo{year}{2012})
  \bibinfo{pages}{020502}.
\bibitem[{Williams et~al.(2012)Williams, Bestwick, Gallagher, Hong, Cui,
  Bleich, Analytis, Fisher, and Goldhaber-Gordon}]{2012arXiv1202.2323W}
\bibinfo{author}{J.~R. Williams}, \bibinfo{author}{A.~J. Bestwick},
  \bibinfo{author}{P.~Gallagher}, \bibinfo{author}{S.~S. Hong},
  \bibinfo{author}{Y.~Cui}, \bibinfo{author}{A.~S. Bleich},
  \bibinfo{author}{J.~G. Analytis}, \bibinfo{author}{I.~R. Fisher},
  \bibinfo{author}{D.~Goldhaber-Gordon}, \bibinfo{journal}{Phys. Rev. Lett.}
  \bibinfo{volume}{109} (\bibinfo{year}{2012}) \bibinfo{pages}{056803}.
\bibitem[{Mourik et~al.(2012)Mourik, Zuo, Frolov, Plissard, Bakkers, and
  Kouwenhoven}]{nature}
\bibinfo{author}{V.~Mourik}, \bibinfo{author}{K.~Zuo}, \bibinfo{author}{S.~M.
  Frolov}, \bibinfo{author}{S.~R. Plissard}, \bibinfo{author}{E.~P. A.~M.
  Bakkers}, \bibinfo{author}{L.~P. Kouwenhoven}, \bibinfo{journal}{Science}
  \bibinfo{volume}{336} (\bibinfo{year}{2012}) \bibinfo{pages}{1003}.
\bibitem[{Duckheim and Brouwer(2011)}]{PhysRevB.83.054513}
\bibinfo{author}{M.~Duckheim}, \bibinfo{author}{P.~W. Brouwer},
  \bibinfo{journal}{Phys. Rev. B} \bibinfo{volume}{83} (\bibinfo{year}{2011})
  \bibinfo{pages}{054513}.
\bibitem[{Chung et~al.(2011)Chung, Zhang, Qi, and Zhang}]{PhysRevB.84.060510}
\bibinfo{author}{S.~B. Chung}, \bibinfo{author}{H.-J. Zhang},
  \bibinfo{author}{X.-L. Qi}, \bibinfo{author}{S.-C. Zhang},
  \bibinfo{journal}{Phys. Rev. B} \bibinfo{volume}{84} (\bibinfo{year}{2011})
  \bibinfo{pages}{060510}.
\bibitem[{Oreg et~al.(2010)Oreg, Refael, and von
  Oppen}]{PhysRevLett.105.177002}
\bibinfo{author}{Y.~Oreg}, \bibinfo{author}{G.~Refael}, \bibinfo{author}{F.~von
  Oppen}, \bibinfo{journal}{Phys. Rev. Lett.} \bibinfo{volume}{105}
  (\bibinfo{year}{2010}) \bibinfo{pages}{177002}.
\bibitem[{Gangadharaiah et~al.(2011)Gangadharaiah, Braunecker, Simon, and
  Loss}]{PhysRevLett.107.036801}
\bibinfo{author}{S.~Gangadharaiah}, \bibinfo{author}{B.~Braunecker},
  \bibinfo{author}{P.~Simon}, \bibinfo{author}{D.~Loss},
  \bibinfo{journal}{Phys. Rev. Lett.} \bibinfo{volume}{107}
  (\bibinfo{year}{2011}) \bibinfo{pages}{036801}.
\bibitem[{Lutchyn et~al.(2010)Lutchyn, Sau, and
  Das~Sarma}]{PhysRevLett.105.077001}
\bibinfo{author}{R.~M. Lutchyn}, \bibinfo{author}{J.~D. Sau},
  \bibinfo{author}{S.~Das~Sarma}, \bibinfo{journal}{Phys. Rev. Lett.}
  \bibinfo{volume}{105} (\bibinfo{year}{2010}) \bibinfo{pages}{077001}.
\bibitem[{Bolech and Giamarchi(2004)}]{PhysRevLett.92.127001}
\bibinfo{author}{C.~J. Bolech}, \bibinfo{author}{T.~Giamarchi},
  \bibinfo{journal}{Phys. Rev. Lett.} \bibinfo{volume}{92}
  (\bibinfo{year}{2004}) \bibinfo{pages}{127001}.
\bibitem[{Bocquet et~al.(2000)Bocquet, Serban, and Zirnbauer}]{Bocquet2000628}
\bibinfo{author}{M.~Bocquet}, \bibinfo{author}{D.~Serban},
  \bibinfo{author}{M.~R. Zirnbauer}, \bibinfo{journal}{Nuclear Physics B}
  \bibinfo{volume}{578} (\bibinfo{year}{2000}) \bibinfo{pages}{628 -- 680}.
\bibitem[{Mudry et~al.(1999)Mudry, Brouwer, and Furusaki}]{PhysRevB.59.13221}
\bibinfo{author}{C.~Mudry}, \bibinfo{author}{P.~W. Brouwer},
  \bibinfo{author}{A.~Furusaki}, \bibinfo{journal}{Phys. Rev. B}
  \bibinfo{volume}{59} (\bibinfo{year}{1999}) \bibinfo{pages}{13221--13234}.
\bibitem[{Mudry et~al.(2000)Mudry, Brouwer, and Furusaki}]{PhysRevB.62.8249}
\bibinfo{author}{C.~Mudry}, \bibinfo{author}{P.~W. Brouwer},
  \bibinfo{author}{A.~Furusaki}, \bibinfo{journal}{Phys. Rev. B}
  \bibinfo{volume}{62} (\bibinfo{year}{2000}) \bibinfo{pages}{8249--8268}.
\bibitem[{Motrunich et~al.(2001)Motrunich, Damle, and
  Huse}]{PhysRevB.63.224204}
\bibinfo{author}{O.~Motrunich}, \bibinfo{author}{K.~Damle},
  \bibinfo{author}{D.~A. Huse}, \bibinfo{journal}{Phys. Rev. B}
  \bibinfo{volume}{63} (\bibinfo{year}{2001}) \bibinfo{pages}{224204}.
\bibitem[{Gruzberg et~al.(2005)Gruzberg, Read, and
  Vishveshwara}]{PhysRevB.71.245124}
\bibinfo{author}{I.~A. Gruzberg}, \bibinfo{author}{N.~Read},
  \bibinfo{author}{S.~Vishveshwara}, \bibinfo{journal}{Phys. Rev. B}
  \bibinfo{volume}{71} (\bibinfo{year}{2005}) \bibinfo{pages}{245124}.
\bibitem[{Muzykantskii and Khmelnitskii(1994)}]{PhysRevB.50.3982}
\bibinfo{author}{B.~A. Muzykantskii}, \bibinfo{author}{D.~E. Khmelnitskii},
  \bibinfo{journal}{Phys. Rev. B} \bibinfo{volume}{50} (\bibinfo{year}{1994})
  \bibinfo{pages}{3982--3987}. 
\bibitem[{Soller et~al.(2012)Soller, Hofstetter, Csonka, Yeyati,
  Sch\"onenberger, and Komnik}]{fdots}
\bibinfo{author}{H.~Soller}, \bibinfo{author}{L.~Hofstetter},
  \bibinfo{author}{S.~Csonka}, \bibinfo{author}{A.~L. Yeyati},
  \bibinfo{author}{C.~Sch\"onenberger}, \bibinfo{author}{A.~Komnik},
  \bibinfo{journal}{Phys. Rev. B} \bibinfo{volume}{85} (\bibinfo{year}{2012})
  \bibinfo{pages}{174512}.
\bibitem[{Cohen et~al.(1962)Cohen, Falicov, and Phillips}]{PhysRevLett.8.316}
\bibinfo{author}{M.~H. Cohen}, \bibinfo{author}{L.~M. Falicov},
  \bibinfo{author}{J.~C. Phillips}, \bibinfo{journal}{Phys. Rev. Lett.}
  \bibinfo{volume}{8} (\bibinfo{year}{1962}) \bibinfo{pages}{316--318}.
\bibitem[{Zawadowski(1967)}]{PhysRev.163.341}
\bibinfo{author}{A.~Zawadowski}, \bibinfo{journal}{Phys. Rev.}
  \bibinfo{volume}{163} (\bibinfo{year}{1967}) \bibinfo{pages}{341}.
\bibitem[{H\"ubler et~al.(2012)H\"ubler, Wolf, Scherer, Wang, Beckmann, and
  v.~L\"ohneysen}]{2010arXiv1012.3867H}
\bibinfo{author}{F.~H\"ubler}, \bibinfo{author}{M.~J. Wolf},
  \bibinfo{author}{T.~Scherer}, \bibinfo{author}{D.~Wang},
  \bibinfo{author}{D.~Beckmann}, \bibinfo{author}{H.~v.~L\"ohneysen},
  \bibinfo{journal}{Phys. Rev. Lett.} \bibinfo{volume}{109}
  (\bibinfo{year}{2012}) \bibinfo{pages}{087004}.
\bibitem[{Zhao et~al.(2004)Zhao, L\"ofwander, and Sauls}]{PhysRevB.70.134510}
\bibinfo{author}{E.~Zhao}, \bibinfo{author}{T.~L\"ofwander},
  \bibinfo{author}{J.~A. Sauls}, \bibinfo{journal}{Phys. Rev. B}
  \bibinfo{volume}{70} (\bibinfo{year}{2004}) \bibinfo{pages}{134510}.
\bibitem[{Levitov and Reznikov(2004)}]{PhysRevB.70.115305}
\bibinfo{author}{L.~S. Levitov}, \bibinfo{author}{M.~Reznikov},
  \bibinfo{journal}{Phys. Rev. B} \bibinfo{volume}{70} (\bibinfo{year}{2004})
  \bibinfo{pages}{115305}.
\bibitem[{Gogolin and Komnik(2006)}]{PhysRevB.73.195301}
\bibinfo{author}{A.~O. Gogolin}, \bibinfo{author}{A.~Komnik},
  \bibinfo{journal}{Phys. Rev. B} \bibinfo{volume}{73} (\bibinfo{year}{2006})
  \bibinfo{pages}{195301}.
\bibitem[{Cuevas et~al.(1996)Cuevas, Mart\'\i{}n-Rodero, and
  Yeyati}]{PhysRevB.54.7366}
\bibinfo{author}{J.~C. Cuevas}, \bibinfo{author}{A.~Mart\'\i{}n-Rodero},
  \bibinfo{author}{A.~L. Yeyati}, \bibinfo{journal}{Phys. Rev. B}
  \bibinfo{volume}{54} (\bibinfo{year}{1996}) \bibinfo{pages}{7366--7379}.
\bibitem[{Tokuyasu et~al.(1988)Tokuyasu, Sauls, and Rainer}]{PhysRevB.38.8823}
\bibinfo{author}{T.~Tokuyasu}, \bibinfo{author}{J.~A. Sauls},
  \bibinfo{author}{D.~Rainer}, \bibinfo{journal}{Phys. Rev. B}
  \bibinfo{volume}{38} (\bibinfo{year}{1988}) \bibinfo{pages}{8823--8833}.
\bibitem[{Chevallier et~al.(2013)Chevallier, Simon, and
  Bena}]{2012arXiv1203.2643C}
\bibinfo{author}{D.~Chevallier}, \bibinfo{author}{P.~Simon},
  \bibinfo{author}{C.~Bena}, \bibinfo{journal}{Phys. Rev. B}
  \bibinfo{volume}{88} (\bibinfo{year}{2013}) \bibinfo{pages}{165401}.
\bibitem[{Tanaka, Kashiwaya (1995)Tanaka and Kashiwaya}]{PhysRevLett.74.3451}
\bibinfo{author}{Y.~Tanaka}, \bibinfo{author}{S.~Kashiwaya}, \bibinfo{journal}{Phys. Rev. Lett.}
  \bibinfo{volume}{74} (\bibinfo{year}{1995}) \bibinfo{pages}{3451}.
\bibitem[{Tinkham(1996)}]{tinkham}
\bibinfo{author}{M.~Tinkham}, \bibinfo{title}{Introduction to
  Superconductivity}, \bibinfo{publisher}{McGraw Hill}, \bibinfo{year}{1996}.
\bibitem[{Beenakker(1992)}]{PhysRevB.46.12841}
\bibinfo{author}{C.~W.~J. Beenakker}, \bibinfo{journal}{Phys. Rev. B}
  \bibinfo{volume}{46} (\bibinfo{year}{1992}) \bibinfo{pages}{12841--12844}.
\bibitem[{Gogolin and Komnik(2006)}]{PhysRevLett.97.016602}
\bibinfo{author}{A.~O. Gogolin}, \bibinfo{author}{A.~Komnik},
  \bibinfo{journal}{Phys. Rev. Lett.} \bibinfo{volume}{97}
  (\bibinfo{year}{2006}) \bibinfo{pages}{016602}.
\bibitem[{Soller and Komnik(2011)}]{Soller2011425}
\bibinfo{author}{H.~Soller}, \bibinfo{author}{A.~Komnik},
  \bibinfo{journal}{Physica E} \bibinfo{volume}{44} (\bibinfo{year}{2011})
  \bibinfo{pages}{425}.
\bibitem[{M\'elin(2004)}]{melin-2004-39}
\bibinfo{author}{R.~M\'elin}, \bibinfo{journal}{Eur. Phys. J. B}
  \bibinfo{volume}{39} (\bibinfo{year}{2004}) \bibinfo{pages}{249}.
\bibitem[{Cottet et~al.(2008)Cottet, Dou\ifmmode~\mbox{\c{c}}\else
  \c{c}\fi{}ot, and Belzig}]{PhysRevLett.101.257001}
\bibinfo{author}{A.~Cottet}, \bibinfo{author}{B.~Dou\ifmmode~\mbox{\c{c}}\else
  \c{c}\fi{}ot}, \bibinfo{author}{W.~Belzig}, \bibinfo{journal}{Phys. Rev.
  Lett.} \bibinfo{volume}{101} (\bibinfo{year}{2008}) \bibinfo{pages}{257001}.
\bibitem[{Cottet and Belzig(2008)}]{PhysRevB.77.064517}
\bibinfo{author}{A.~Cottet}, \bibinfo{author}{W.~Belzig},
  \bibinfo{journal}{Phys. Rev. B} \bibinfo{volume}{77} (\bibinfo{year}{2008})
  \bibinfo{pages}{064517}.
\bibitem[{Piano et~al.(2011)Piano, Grein, Mellor, V\'yborn\'y, Campion, Wang,
  Eschrig, and Gallagher}]{PhysRevB.83.081305}
\bibinfo{author}{S.~Piano}, \bibinfo{author}{R.~Grein}, \bibinfo{author}{C.~J.
  Mellor}, \bibinfo{author}{K.~V\'yborn\'y}, \bibinfo{author}{R.~Campion},
  \bibinfo{author}{M.~Wang}, \bibinfo{author}{M.~Eschrig},
  \bibinfo{author}{B.~L. Gallagher}, \bibinfo{journal}{Phys. Rev. B}
  \bibinfo{volume}{83} (\bibinfo{year}{2011}) \bibinfo{pages}{081305}.
\bibitem[{Grein et~al.(2010)Grein, L\"ofwander, Metalidis, and
  Eschrig}]{PhysRevB.81.094508}
\bibinfo{author}{R.~Grein}, \bibinfo{author}{T.~L\"ofwander},
  \bibinfo{author}{G.~Metalidis}, \bibinfo{author}{M.~Eschrig},
  \bibinfo{journal}{Phys. Rev. B} \bibinfo{volume}{81} (\bibinfo{year}{2010})
  \bibinfo{pages}{094508}.
\bibitem[{Grein et~al.(2009)Grein, Eschrig, Metalidis, and
  Sch\"on}]{PhysRevLett.102.227005}
\bibinfo{author}{R.~Grein}, \bibinfo{author}{M.~Eschrig},
  \bibinfo{author}{G.~Metalidis}, \bibinfo{author}{G.~Sch\"on},
  \bibinfo{journal}{Phys. Rev. Lett.} \bibinfo{volume}{102}
  (\bibinfo{year}{2009}) \bibinfo{pages}{227005}.
\bibitem[{Colci et~al.(2012)Colci, Sun, Shah, Vishveshwara, and
  Van~Harlingen}]{PhysRevB.85.180512}
\bibinfo{author}{M.~Colci}, \bibinfo{author}{K.~Sun},
  \bibinfo{author}{N.~Shah}, \bibinfo{author}{S.~Vishveshwara},
  \bibinfo{author}{D.~J. Van~Harlingen}, \bibinfo{journal}{Phys. Rev. B}
  \bibinfo{volume}{85} (\bibinfo{year}{2012}) \bibinfo{pages}{180512}.
\bibitem[{Buzdin(2005)}]{RevModPhys.77.935}
\bibinfo{author}{A.~I. Buzdin}, \bibinfo{journal}{Rev. Mod. Phys.}
  \bibinfo{volume}{77} (\bibinfo{year}{2005}) \bibinfo{pages}{935--976}.
\bibitem[{Volkov et~al.(1995)Volkov, Magnée, van Wees, and
  Klapwijk}]{Volkov1995261}
\bibinfo{author}{A.~Volkov}, \bibinfo{author}{P.~Magnée},
  \bibinfo{author}{B.~van Wees}, \bibinfo{author}{T.~Klapwijk},
  \bibinfo{journal}{Physica C: Superconductivity} \bibinfo{volume}{242}
  (\bibinfo{year}{1995}) \bibinfo{pages}{261 -- 266}.
\bibitem[{Miranda et~al.(1982)Miranda, Yndur\'ain, Chandesris, Lecante, and
  Petroff}]{PhysRevB.25.527}
\bibinfo{author}{R.~Miranda}, \bibinfo{author}{F.~Yndur\'ain},
  \bibinfo{author}{D.~Chandesris}, \bibinfo{author}{J.~Lecante},
  \bibinfo{author}{Y.~Petroff}, \bibinfo{journal}{Phys. Rev. B}
  \bibinfo{volume}{25} (\bibinfo{year}{1982}) \bibinfo{pages}{527--530}.
\end{thebibliography}
\end{document}